\documentclass{vgtc}                          

\usepackage{times}                     

\usepackage{tabu} 
\usepackage{graphicx}
\usepackage{amsmath}

\usepackage{booktabs}                  
\usepackage{lipsum}                    
\usepackage{mwe}                       
\usepackage{textcomp}
\usepackage{diagbox}
\usepackage{booktabs}
\usepackage{colortbl}
\usepackage{multirow}
\usepackage{tabularx}
\usepackage{tabularray}
\usepackage{booktabs}
\usepackage{siunitx}
\usepackage{soul}
\usepackage{wrapfig}
\usepackage[caption=false]{subfig}
\usepackage[font=footnotesize,labelfont=bf]{caption}
\usepackage[normalem]{ulem}
\usepackage{tikz}
\usepackage{xcolor}
\usepackage[dvipsnames]{xcolor}
\usepackage{enumitem}
\usepackage{mathptmx}  
\usepackage{comment}
\usepackage{xspace}

\newcommand{\circled}[2][blue!30]{%
  \tikz[baseline=(char.base)]{
    \node[
      shape=circle,
      draw=black,
      fill=#1,
      inner sep=1pt
    ] (char) {#2};
  }%
}

\newcommand{\circledblack}[2][black]{%
  \circled[#1]{\textcolor{white}{#2}}%
}

\newcommand{\NumberCircle}[2][black]{%
  \tikz[baseline=-0.5ex]
    \node[draw, circle, fill=#1, inner sep=0pt, minimum size=2.0ex, line width =0.4pt, text=black] {#2};%
}

\newcommand{\ProgressBar}[3]{%
  \tikz[baseline=-0.6ex]{
    \draw[draw=black, line width=0.4pt] (0,0) rectangle (#2,#3);
    \fill[black] (0,0) rectangle ({#1*#2},#3);
  }%
}

\newcommand{\circledwhite}[1]{%
  \circled[white]{#1}%
}

\newcommand{\XRPRISM}{\texttt{XR-PRISM}\xspace}

\onlineid{0}

\vgtccategory{Research}

\vgtcinsertpkg

\title{XR-PRISM: Data-Driven Privacy and Risk Impact Scoring Metric for Extended Reality in Healthcare}

\author{Nafisa Anjum\thanks{e-mail: nanjum1@students.kennesaw.edu}\\ %
        \scriptsize Dept. of Computer Science\\ Kennesaw State University %
\and M. Rasel Mahmud\thanks{e-mail: m.raselmahmud1@gmail.com}\\ %
     \scriptsize Assistant Professor of Computer Science\\ Kennesaw State University \\ %
     \parbox{.8in}{\scriptsize \centering}}

\abstract{
Extended Reality (XR) technologies are transforming healthcare, enabling immersive training, remote consultation, and patient rehabilitation; yet their rich sensor and data pipelines introduce novel privacy and safety vulnerabilities. To date, there is no unifying, quantitative framework to assess these risks. In this paper, we survey 65 peer-reviewed XR security and privacy studies (2017–2024), synthesizing a four-layer (Device, Network, User, Cloud) threat taxonomy and catalog of defenses. Building on this foundation, we propose \XRPRISM, a six-factor, weighted Privacy and Risk Impact Scoring Metric that integrates threat likelihood, system vulnerabilities, attack surface, safety impact, privacy impact, and control effectiveness into a single actionable score. Our analysis uncovers that over 70\% of countermeasures lack standardized risk evaluations and fewer than 15\% of the attacks need high expertise to be launched. \XRPRISM offers practitioners a transparent, data-driven tool for prioritizing and mitigating security and privacy risks in XR healthcare deployments.
} 


\begin{document}


\firstsection{Introduction}

\maketitle
Extended Reality (XR), encompassing Virtual Reality (VR), Augmented Reality (AR), and Mixed Reality (MR), is transforming healthcare by enabling immersive training, remote consultation, patient rehabilitation, and mental-health therapies. Many VR-based assistive feedback improved balance and gait impairments \cite{9995441,9756779}. Yet, the very high-fidelity motion traces, physiological signals, biometric identifiers, and rich environmental context that make XR so powerful also expose patients and providers to unprecedented security and privacy (S\&P) risks. Moreover, machine learning (ML) and deep learning (DL) components are becoming integral to XR systems—powering gesture recognition, environment mapping, anomaly detection, and personalized therapy. While some works document individual vulnerabilities \cite{giaretta2024security}, there is still no unifying framework that systematically characterizes these threats, and most defenses lack rigorous, standardized risk evaluations or recovery mechanisms once prevention fails. To fill this gap, we conduct a Systematization of Knowledge (SoK) by surveying 65 peer-reviewed XR S\&P studies (2017–2024) and contribute:
\vspace{-0.1in}
\begin{itemize}
    \item \textbf{Four-layer taxonomy}: Deconstructing XR into Device, Network, User, and Cloud layers, and classifying threats and countermeasures within each.
    \vspace{-0.1in}
    \item \textbf{Threat–defense mapping}: Rigorous analysis of representative attack vectors and defenses, highlighting prerequisites, attacker expertise, efficacy, and overhead.
    \item \textbf{XR-PRISM}: A six-factor, weighted Privacy and Risk Impact Scoring Metric that integrates threat likelihood, XR system vulnerabilities, attack surface, safety impact, privacy impact, and control effectiveness into a single actionable score.
\end{itemize}
\vspace{-0.1in}
\section{Related Work}\label{sec:related}
The security domain has examined risk assessment, where risk is frequently quantified in relation to the probability and consequences of a security threat\cite{maclean2017nist}. \textit{Wagner et al.}\cite{wagner2018privacy} proposes a way to measure and depict privacy risk that takes into account a number of variables as well as various scenarios and attacker types. \textit{Wu et al.}\cite{wu2023personal} proposes a framework on disclosed  personal identifiable information. A method for determining a business process's fundamental components and evaluating their security quantitatively was put out by \textit{Bhattacharjee et al.}\cite{bhattacharjee2016quantitative}. To facilitate privacy impact evaluation during the early stages of information system development, \textit{Ahamdian et al.}\cite{ahmadian2018supporting} presented a model-based privacy analysis. But existing research either quantify risks for enterprise workflows or discrete user-provided attributes only. They do not address the unique characteristics of XR for comparable risk scores or support both preventive and post‑compromise evaluations; despite a rapidly growing corpus of work on XR applications in healthcare. This outlines a lack of mechanisms for scoring combined safety and privacy impacts in clinical XR deployments.
\vspace{-0.01in}
\section{Methodology}\label{sec:method}
In this section , we first organize an XR system pipeline into four concentric layer architecture. Subsequently, we conduct a comprehensive investigation evaluating S\&P publications in XR.
\paragraph{\textbf{Architecture.}}Modern XR headsets integrate multiple sensors—Inertial Measurement Units (IMUs), optical trackers, and eye-gaze modules to capture significant data from the user and the environment at high sampling rates\cite{sheng2024review}. We conceptualize XR system pipeline as four concentric layers:  
\vspace{-0.1in}
\begin{enumerate}[label=(\arabic*)]
  \item \textbf{User layer}: Biometric and behavioral signals (gestures, voice commands, EMG, heart rate) captured via wearables and controllers.  
  \vspace{-0.1in}
  \item \textbf{Device layer}: Onboard compute and firmware responsible for sensor fusion, local rendering, and real‐time data preprocessing.  
  \vspace{-0.1in}
  \item \textbf{Network layer}: Encrypted transport channels (Wi-Fi 6E, 5G, BLE) conveying telemetry, video streams, and control commands between headsets and edge or cloud endpoints.
  \vspace{-0.1in}
  \item \textbf{Cloud layer}: Remote compute for analytics, long‐term storage, federated learning, and compliance‐enforced data repositories.  
\end{enumerate} 
\vspace{-0.1in}
In healthcare settings, XR has proven effective for motor rehabilitation (stroke gait training, balance therapy), cognitive therapy (exposure treatment for phobias), and surgical assistance (AR overlays for anatomical guidance), with multiple clinical trials reporting statistically significant improvements in patient outcomes\cite{yang2021utilization}.

\paragraph{\textbf{Systematization.}}
\begin{table}[t]
  \footnotesize
  \setlength{\tabcolsep}{4pt}
  \caption{Keyword Strategies for XR-Healthcare}
  \label{tab:keyword-strategies}
  \begin{tabular}{@{}l p{0.68\columnwidth}@{}}
    \toprule
    \textbf{Groups}      & \textbf{Terms} \\
    \midrule
    XR Modality            & virtual reality \textbf{OR} augmented reality \textbf{OR} mixed reality \textbf{OR} extended reality \\
    Healthcare Context     & healthcare \textbf{OR} tele-rehabilitation \textbf{OR} therapy \\
    Security Focus         & security \textbf{OR} attack \textbf{OR} threat \\
    Privacy Focus          & privacy \textbf{OR} differential privacy \textbf{OR} PHI (Protected Health Information) \\
    Defense Mechanisms     & mitigation \textbf{OR} countermeasure \textbf{OR} access control \textbf{OR} encryption \textbf{OR} obfuscation \\
    Quantitative Metrics   & risk assessment \textbf{OR} CVSS (Common Vulnerability Scoring System\cite{first2019cvss}) \textbf{OR} framework \\
    \bottomrule
  \end{tabular}
  \vspace{-0.2in}
\end{table}
We performed a PRISMA–compliant \cite{fink2019conducting} systematic literature review (SLR) over four databases: \textit{IEEE Xplore, ACM Digital Library, USENIX Proceedings, and PubMed}. Our search combined six keyword groups covering: \circledblack{1} XR Modality, \circledblack{2} Healthcare Context, \circledblack{3} Security Focus, \circledblack{4} Privacy Focus, \circledblack{5} Defense Mechanisms, \circledblack{6} Quantitative Metrics; the details of the terms in each group are included in Table {\ref{tab:keyword-strategies}}. Our initial query (2017–2024) returned 207 records. After removing 91 duplicates, title/abstract screening, and full‐text assessment in teams, we select inclusion criteria—peer‐reviewed studies that \circledwhite{1} analyze XR threats or \circledwhite{2} propose defenses with experimentation and metrics. Studies that explicitly analyze threats or propose defenses—attacks, side-channels, authentication, data protection or recovery mechanisms in XR environments that can also be applied to the healthcare industry were considered as the prime focus. Finally, we retained 65 publications now available in our GitHub \footnote{\url{https://github.com/User32-blip/SoK-XR-in-Healthcare}}.
For each paper, we extracted:
\vspace{-0.1in}
\begin{itemize}[noitemsep]
  \item \emph{Attack attributes}: XR layer targeted, attack vector and component, adversary prerequisites (hardware/software access), required expertise of attacker, and impact or the degree to which an attack exposes sensitive patient data based on the performance.
  \item \emph{Defense attributes}: Defense group, mitigation vector, overhead, maintainability, efficacy based on evaluation reported, and defense stage (prevention ($P$), detection ($D$), recovery($R$)).  
\end{itemize}

These attributes were aggregated into our four‐layer threat taxonomy.  We then applied thematic analysis to identify gaps such as underexplored recovery mechanisms and inconsistent risk reporting and distilled the six core factors for our \XRPRISM quantitative framework detailed in Sec.\ref{sec:xr-prism}.
\section{Overview of Threats and Defenses}
\paragraph{\textbf{Target: Device Layer.}}
In shared virtual environments, \textit{Yang et al.}\cite{yang2024can} create and execute a novel class of keystroke inference attacks that allow an attacker (VR user) to retrieve material typed by another VR user by looking at their avatar. \textit{Slocum et al.}\cite{slocum2023going} demonstrate how an attacker can easily extract stream head tracking data from an AR/VR device, segment it, and categorize it to get private text data.
\paragraph{\textbf{Target: User Layer.}}
\textit{Gopal et al.}\cite{gopal2023hidden} introduces Hidden Reality \textit{(HR Model)}, a video-based side-channel attack that demonstrates how, even if the virtual screen in VR devices is not directly visible to adversaries, indirect observations could be used to acquire the user's personal data.
\paragraph{\textbf{Target: Network Layer.}}
\textit{Arafat et al.}\cite{al2021vr} introduced \textit{VR-Spy}, a brand-new human activity-based side-channel attack that infers text inputs from virtual reality devices. The fundamental concept of VR-Spy is that each virtual keystroke has a distinct gesture pattern in the CSI waveforms based on the side-channel information of fine-granular hand movements.
\paragraph{\textbf{Target: Cloud Layer.}}
\textit{Tseng et al.}\cite{tseng2022dark} exploited the fact that VR platforms typically trust any code running in a VR app with direct, unmediated access to the user’s body-tracking streams and environment model. 
\begin{figure}[t]
  \centering
  \includegraphics[width=0.85\columnwidth]{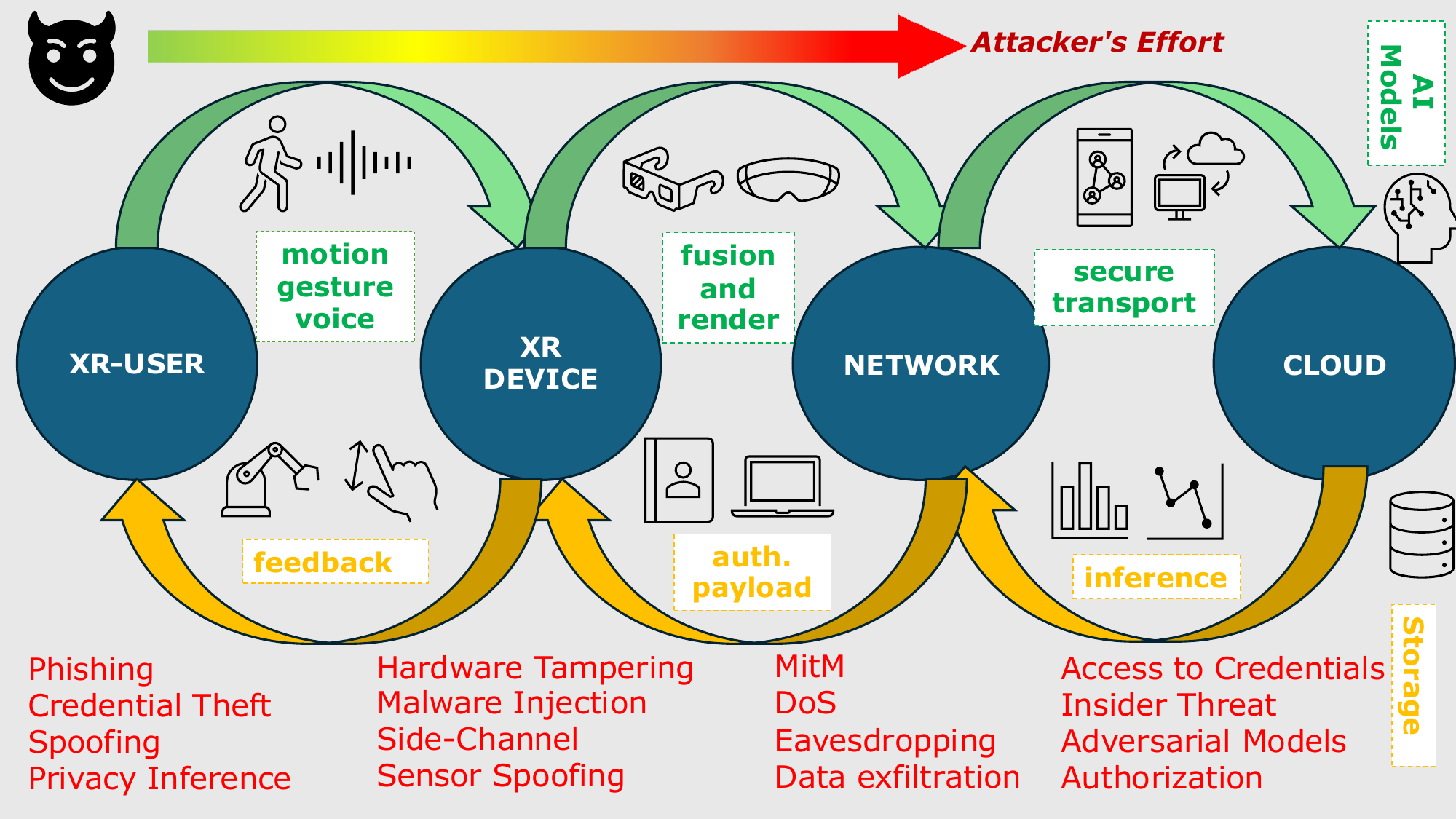}
  \caption{Threat model of an XR system.}
  \label{fig:xr-threat-model}
  \vspace{-0.2in}
\end{figure}

\begin{table*}[t]
  \centering
  \footnotesize
  \caption{An overview of XR Attack Methods (\NumberCircle[green]{ } Low \NumberCircle[yellow]{ } Medium \NumberCircle[red]{ } High)}
  \label{tab:attack-taxonomy}
  \begin{tabular}{@{}lllllccl@{}}
    \toprule
    \multirow{2}{*}{\textbf{Approach}} 
      & \multirow{2}{*}{\textbf{Layer}} 
       & \multirow{2}{*}{\textbf{Technique}} 
      & \multicolumn{2}{c}{\textbf{Dimension}} 
      & \multicolumn{2}{c}{\textbf{Complexity}}  
      & \multirow{2}{*}{\textbf{Impact}} \\
    \cmidrule(lr){4-5} \cmidrule(lr){6-7}
      &  &  &  \textbf{Attack Vector} & \textbf{Component} 
      & \textbf{Requisite} & \textbf{Expertise} &   \\
    \midrule
   TyPose \cite{slocum2023going} &Device &Keystroke Inference &IMU/head tracking &On-device motion sensors &\NumberCircle[red]{ } &\NumberCircle[yellow]{ } &\ProgressBar{0.82}{3em}{0.8ex} \\
    

    Hand gesture \cite{gopal2023hidden} &\multirow{1}{*}{User} &Exploit typing gesture &Side-Channel &Video segment &\NumberCircle[yellow]{ } &\NumberCircle[red]{ } &\ProgressBar{0.75}{3em}{0.8ex}\\
    VR-Spy \cite{al2021vr} &Network &Wireless Sniffing &Virtual Keystrokes &CSI data &\NumberCircle[red]{ } &\NumberCircle[red]{ } &\ProgressBar{0.65}{3em}{0.8ex}\\
     Run Malicious Code \cite{tseng2022dark} &\multirow{1}{*}{Cloud} &Remote code execution &Software &Cloud storage &\NumberCircle[red]{ } &\NumberCircle[red]{ } &\ProgressBar{0.8}{3em}{0.8ex}\\
    \bottomrule
  \end{tabular}
\end{table*}
\vspace{-0.1in}

\begin{table*}[t]
  \centering
  \footnotesize
  \caption{An overview of XR Defense Approaches (\NumberCircle[green]{ } Low \NumberCircle[yellow]{ } Medium \NumberCircle[red]{ } High)}
  \label{tab:defense-approaches}
  \begin{tabular}{@{} lll l cc l c @{}}
    \toprule
    \multirow{2}{*}{\textbf{Approach}}
      & \multirow{2}{*}{\textbf{Group}}
      & \multirow{2}{*}{\textbf{Mitigation}}
       & \multirow{2}{*}{\textbf{Defense Vector}}
      & \multicolumn{2}{c}{\textbf{Deployability}}
      & \multicolumn{2}{c}{\textbf{Robustness}}
      \\
    \cmidrule(lr){5-6} \cmidrule(lr){7-8}
      &  &  &  & \textbf{Trade-off} & \textbf{Maintenance} & \textbf{Efficacy}  &  \textbf{Stage}\\
      
    \midrule
    Keystroke inference \cite{yang2024can}
      & \multirow{1}{*}{Data Obfuscation} & Limit access to telemetry & Hand tracking API     
      & \NumberCircle[green]{ }             & \NumberCircle[green]{ }         
      &\ProgressBar{0.55}{3em}{0.8ex}            & P\\

     Biometric Auth.\cite{wang2021nod} &\multirow{1}{*}{Authentication} &Head-neck motion &IMU telemetry &\NumberCircle[green]{ } &\NumberCircle[yellow]{ } &\ProgressBar{0.9}{3em}{0.8ex} &P\\
       
      ShareAR\cite{ruth2019secure} &\multirow{1}{*}{Access Control} &Physical-world controls &App-level APIs &\NumberCircle[green]{ }  &\NumberCircle[yellow]{ }  &\ProgressBar{0.8}{3em}{0.8ex} &P\\
     \bottomrule
  \end{tabular}
  \vspace{-0.1in}
\end{table*}
\section{XR-PRISM}\label{sec:xr-prism}
Our findings show that most XR side-channel and inference attacks require only minimal privileges (scores of 1–2) and modest expertise—few demand specialized hardware or deep reverse-engineering. Healthcare XR deployments blend rich sensory inputs, real‐time rendering, haptic feedback, and sensitive biometric streams, creating intertwined security and privacy exposures. To quantify and prioritize these exposures, we extend a \textit{Multi‐Criteria Decision Analysis (MCDA)}\cite{linkov2011multi} based scoring framework to jointly assess both security and privacy risks. This approach consists of two main steps:\circledblack{1}
Scaling key parameters as per threat characteristics and \circledblack{2}Calculating the \textit{RiskScore} for taking immediate mitigation action. These key elements together form the \XRPRISM (XR-Privacy and Risk Impact Scoring Metric).
\begin{table}[ht]
  \centering
  \footnotesize
  \caption{Risk Assessment Model Parameters}
  \label{tab:risk-parameters}
  \begin{tabular}{@{}llc@{}}
    \toprule
    \textbf{Risk Factor}    & \textbf{Description}   & \textbf{Weight($W$)} \\
    \midrule
    \textbf{Threat Likelihood} ($L$)   & Probability of attack        & 0.15 \\
    \textbf{System Vulnerabilities} ($V$) & Known XR platform flaws             & 0.15 \\
    \textbf{Attack Surface} ($A$)      & Exposure of interfaces        & 0.10 \\
    \textbf{Safety Impact} ($I_s$)     & Patient-harm         & 0.30 \\
    \textbf{Privacy Impact} ($I_p$)    & User identification           & 0.20 \\
    \textbf{Control Effectiveness} ($C$) & Strength of auth, encryption     & 0.10 \\
    \bottomrule
  \end{tabular}
  \vspace{-0.2in}
\end{table}
\paragraph{\textbf{Key Risk Factors for Weighting}}
The proposed structure in \cite{ganin2020multicriteria} is intended to evaluate a cyber system’s risk using threats, vulnerabilities and consequences as the most significant criteria in order to choose the best remedial strategy. Expanding on their idea and the CVSS scoring system, the scoring mechanism developed here scores each risk factor from 1 (low risk) to 10 (high
risk). We began by assigning weights to the six risk criteria, namely, \circledblack{1}Threat Likelihood, \circledblack{2}System Vulnerabilities, \circledblack{3}Attack Surface, \circledblack{4}Safety Impact, \circledblack{5} Privacy Impact and \circledblack{6} Control Effectiveness; shown in Table \ref{tab:risk-parameters}. These weights would be obtained from XR security specialists using established procedures \cite{buede2024engineering} in an empirical implementation of this paradigm, depending on the attributes of the XR healthcare system. We treat these weights as initial heuristics; we plan a Delphi-style expert elicitation to empirically calibrate them. The values of the scale have been interpreted as per NIST SP 800-30 guidelines \cite{nist80030} but are susceptible to change on the basis of $L$.

\paragraph{\textbf{Formula and Scoring Interpretation}} Subsequent to defining and quantifying the parameters, the overall risk score is calculated using a weighted sum in Eq.\ref{eq:riskscore}. Control Effectiveness ($C$) is subtracted from 10, which is the highest score, because stronger controls reduce risk.

We score six factors on a 1–10 scale and compute a single \emph{RiskScore} as a weighted sum:
\begin{equation}\label{eq:riskscore}
\mathit{RiskScore} = L\,W_{L} + V\,W_{V} + A\,W_{A} + I_{s}\,W_{I_{s}} + I_{p}\,W_{I_{p}} + (10 - C)\,W_{C}
\end{equation}

where:
\begin{itemize}
\vspace{-0.1in}
  \item $L$ (\emph{Threat Likelihood}): probability of an attack, informed by incident data and exploitability indices.
  \vspace{-0.1in}
  \item $V$ (\emph{System Vulnerabilities}): count and severity of known flaws in firmware, runtime, and architecture.\vspace{-0.1in}
  \item $A$ (\emph{Attack Surface}): number and exposure level of sensors, APIs, and network links\vspace{-0.1in}
  \item $I_s$ (\emph{Safety Impact}): potential for patient harm (haptics, motion-sickness)\vspace{-0.1in} 
  \item $I_p$ (\emph{Privacy Impact}): severity of PHI leakage or behavioral profiling\vspace{-0.1in} 
  \item $C$ (\emph{Control Effectiveness}): strength of authentication, encryption, and session isolation (higher $ C$ is more effective).\vspace{-0.1in}
\end{itemize}

We choose weights to reflect the paramount importance of patient safety and data confidentiality:
\begin{equation}
\begin{aligned}
W_L &= 0.15,\quad W_V = 0.15,\quad W_A = 0.10,\\
W_{I_s} &= 0.30,\quad W_{I_p} = 0.20,\quad W_C = 0.10,
\end{aligned}
\end{equation}
with $\sum W=1.00$.

Using the formula, the risk score that is calculated is assigned Risk Levels from Low to Critical as per Table \ref{tab:risk-interpretation}. From the risk level, the required mitigation priority and appropriate action to be undertaken for the threat can be determined. An organization prioritizes outcomes and controls that can manage the risks with the most negative impacts and that are most cost-effective for their risk management results by using the
principles outlined in NIST SP 800-53: Security and Privacy Controls for Information Systems and Organizations \cite{force2013security}. \XRPRISM extends beyond CVSS by explicitly folding in safety and privacy impacts—critical in healthcare XR, via two dedicated factors, \textit{Safety Impact} and \textit{Privacy Impact} each weighted heavily to reflect patient‐harm and PHI leakage concerns. XR-PRISM can also integrate modifiers for privacy controls such as differential privacy noise budgets\cite{david2021privacy}, secure multi-party computation, or anonymization thresholds to penalize residual inference risk.
\begin{table}[ht]
  \centering
  \footnotesize
  \caption{Risk Score Interpretation}
  \label{tab:risk-interpretation}
  \begin{tabular}{@{}ccc@{}}
    \toprule
    \textbf{RiskScore} & \textbf{Risk Level}  & \textbf{Mitigation Action}      \\
    \midrule
    $1$–$3$            & \textbf{Low}       & Monitor routinely;no immediate change  \\
    $4$–$6$            & \textbf{Moderate}  & Deploy preventive controls\\
    $7$–$8$            & \textbf{High}      & Immediate mitigation; elevate priority  \\
    $9$–$10$           & \textbf{Critical}  & Emergency response; consider system shutdown \\
    \bottomrule
  \end{tabular}
  \vspace{-0.1in}
\end{table}

\paragraph{Example.}
A tele‐therapy VR system suffers a motion‐replay attack that risks both user disorientation and PHI inference. Experts rate:
\begin{equation}\label{eq:values}
    L=6,\;V=5,\;A=7,\;I_s=8,\;I_p=9,\;C=4.
\end{equation}
\vspace{-0.2in}
\begin{equation}\label{eq:example}
  \resizebox{\columnwidth}{!}{$
    \mathit{RiskScore}
      =6\cdot0.15 + 5\cdot0.15 + 7\cdot0.10
       + 8\cdot0.30 + 9\cdot0.20 + (10-4)\cdot0.10
      =6.5
  $}
\end{equation}
placing it in the **High** tier.  We therefore recommend urgent deployment of signed telemetry, anomaly detection at the edge, and end‐to‐end encryption of all biometric streams.

\section{Research Gaps}\label{sec:challenges}

By examining the information XR S\&P publications through extracting information outlined in Sec.\ref{sec:method}, we now concentrate on determining research gaps and suggestions for further research.
\paragraph{\textbf{Low Prerequisites for Most Attacks.}}
A data analysis of papers similar to Table \ref{tab:attack-taxonomy} illustrates that the majority of documented attack methods demand minimal prerequisites; only a handful \cite{slocum2023going,zhang2023s} reach a $High$ prerequisite level. This skew toward low-barrier exploits suggests that XR systems are broadly exposed to attacks by relatively unsophisticated adversaries, underscoring the urgent need to elevate baseline security measures. 
\paragraph{\textbf{ML/DL Threat Surfaces.}}
Modern XR systems increasingly embed machine learning for gesture recognition, anomaly detection, and personalization thus opening new vulnerabilities. Examples include adversarial perturbations that mislead model outputs, model-inversion that reconstructs sensitive training data \cite{qayyum2024secure}, and inference attacks that extract behavioral profiles. To score these, XR-PRISM can be expanded to incorporate exploitability metrics from ML robustness benchmarks to gauge how easily adversarial inputs can be generated.
\paragraph{\textbf{Predominance of Preventative Controls.}}
Nearly all surveyed defenses are preventative—aimed at blocking attacks before they occur; yet there is a dearth of mechanisms for detection, forensics, or automated recovery post-compromise. We note that fewer than over 70\% of countermeasures lack standardized risk evaluations, and only 15\% include user-study driven usability assessments. This lack of holistic, end-to-end security frameworks. 
\paragraph{\textbf{Underexplored Cloud-Layer Threats and Defenses.}}
Even though works like the COVID-19 XR-IoMT system \cite{tai2021trustworthy} hinges on a 5G-backed cloud infrastructure to aggregate, process, and secure sensitive medical data, emerging cloud threats such as poisoning federated learning updates to degrade model integrity or side-channel inference on aggregated telemetry are largely unstudied.
\begin{figure}[t]
  \centering
  \includegraphics[width=0.75\columnwidth]{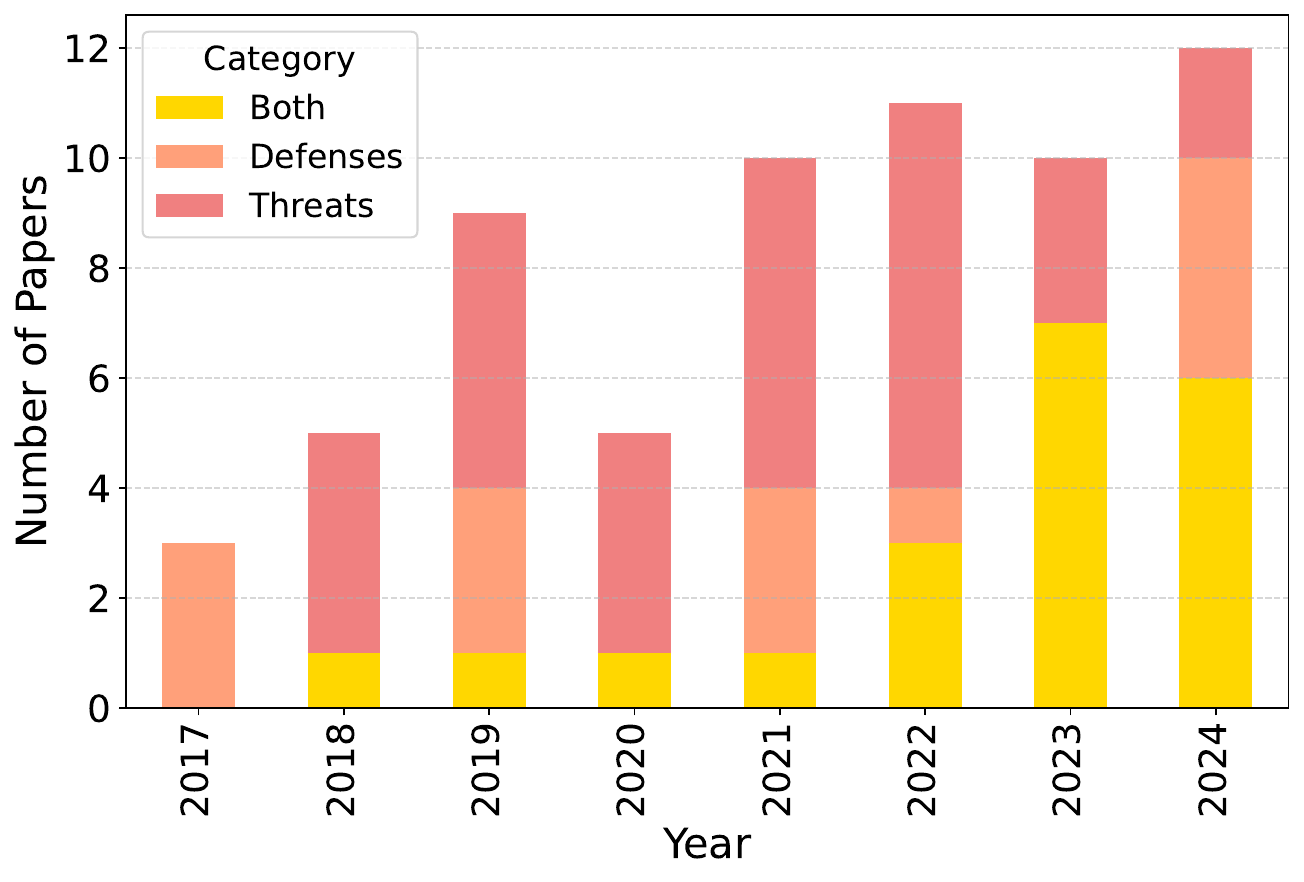}
  \caption{Publications by Year and Category }
  \label{fig:chart}
  \vspace{-0.2in}
\end{figure}

\section{Conclusion}\label{sec:conclusion}
 The delivery of care and possibly the safety and well-being of people may be impacted if an attacker compromises a clinical XR device and tampers with the content of a clinical XR immersive session. In this paper, we provide the first comprehensive systematization of knowledge on XR privacy and security in healthcare, surveying 65 studies and organizing threats and defenses into a four-layer taxonomy; we introduce \XRPRISM, a weighted six-factor risk-scoring framework that unifies into a single actionable metric which can be utilized to interpret the required level of action for threat mitigation. Future work will apply \XRPRISM to real XR healthcare case studies, e.g., VR stroke rehabilitation and AR guided surgery, and validate scores against incident logs and practitioner feedback.
\bibliographystyle{abbrv-doi}

\bibliography{bibliography}
\end{document}